\documentclass{aa}  
\usepackage{graphicx}
\usepackage{natbib}
\usepackage{txfonts}
\usepackage[breaklinks=true]{hyperref}
\usepackage{color}
\usepackage{tikz}
\usetikzlibrary{shapes,arrows}

\newcommand{\reflexstyle}[1]{{\sl #1}}

\begin{document}

\title{Automated data reduction workflows for astronomy}

\subtitle{The ESO Reflex environment}

\titlerunning{Data reduction workflows}

   \author{W. Freudling, 
           M. Romaniello,
           D.M. Bramich,
           P. Ballester,
           V. Forchi,
           C.E. Garc\'ia-Dabl\'o,
           S. Moehler,
          \and
           M.J. Neeser
          }
   \authorrunning{Freudling et al.}

   \institute{European Southern Observatory, \\
              Karl-Schwarzschild-Str. 2, \\
              85748 Garching,  \\
              Germany \\
              \email{wfreudli@eso.org}
             }

   \date{\sl{To appear in Astronomy \& Astrophysics, Volume 559, November 2013, A96}}

\abstract{}{}{}{}{} 
 
\abstract 
{Data from complex modern astronomical instruments often consist of a large
number of different science and calibration files, and their reduction requires
a variety of software tools.  The execution chain of the tools represents a
complex  workflow that needs to be tuned and supervised, often by individual
researchers that are not necessarily experts for any specific instrument.}
{The efficiency of data reduction can be improved by using automatic workflows
to organise data and execute a sequence of data reduction steps.  To realize
such efficiency gains, we designed a system that allows intuitive
representation, execution and modification of the data reduction workflow,  and
has facilities for inspection and interaction with the data.} 
{The European Southern Observatory (ESO) has developed \reflexstyle{Reflex}, an
environment to automate data reduction workflows.  \reflexstyle{Reflex} is
implemented as a package of customized components  for the Kepler workflow
engine. Kepler provides the graphical user interface to
create an executable flowchart-like representation of the data reduction
process. Key features of \reflexstyle{Reflex} are a rule-based data organiser,
infrastructure to re-use results,  thorough book-keeping, data progeny
tracking, interactive user interfaces, and  a novel concept to exploit
information created during  data organisation for the workflow execution.  } 
{ Automated workflows can greatly increase the efficiency of astronomical data
reduction.  In \reflexstyle{Reflex}, workflows can be run non-interactively as
a first step. Subsequent optimization can then be carried out while
transparently re-using all unchanged intermediate products. We found that such
workflows enable the reduction of complex data by non-expert users and
minimizes mistakes due to book-keeping errors.} 
{\reflexstyle{Reflex} includes novel concepts to increase the efficiency of
astronomical data processing.  While \reflexstyle{Reflex} is a specific
implementation of astronomical scientific workflows within the Kepler workflow
engine, the overall design choices and methods can also be applied to other
environments for running automated science workflows.}


\maketitle
%

\section{Introduction}

Astronomical observations produce data streams that  record the signal of
targets and carry associated metadata that include observational parameters and
a host of associated information. Apart from the intended signal, such raw data
include signatures of the atmosphere and the instrument, as well as noise from
various sources. Before any scientific analysis of the data, a process called
``data reduction'' is used to remove the instrumental signature and contaminant
sources  and, for ground based observations, remove atmospheric effects.
Only then, can the signal of the target source be extracted. In
general, data reduction also includes a noise model and error propagation
calculations to estimate uncertainties in the extracted signal. 

In recent years,  astronomical data reduction and analysis has become
increasingly complex.  The data from modern instruments can now comprise dozens
of different data types that include both science and calibration data. For
example, the reduction of data from ESO's  X-Shooter instrument uses almost 100
different data types for its three simultaneously working arms.  The reduction
of such data in general includes a large number of complex high-level
algorithms and methods. The data types and methods are interdependent in a
complex web of relations. It is therefore increasingly difficult for an
individual researcher to understand, execute and optimize a data reduction
cascade for complex instruments. This situation has led to the appearance of
specialized, highly integrated data reduction pipelines that are written by
specialists and can reduce data without supervision
\citep[e.g.][]{bib:wfpcpipe,bib:vltpipe,bib:sdsspipe,bib:theli,bib:pipeenv,bib:robonet}.
{For efficient large scale data reduction, such pipelines often run in
custom-made environments.  For example, ESO  employs a system for quality
control that automatically associates calibration data and processes them as
soon as they arrive from the telescopes.  The results are then stored in its
data archive.  Other examples of such  event driven data reduction environments
are NOAO's High-Performance Pipeline \citep{bib:noao}, STScI's OPUS system
\citep{bib:opus}, and the Astro-WISE pipeline \citep{bib:astrowise}.

Automatic pipelines work best for data from long-term projects that use  stable
instrumentation, aim for a well-defined set of similar targets observed at
similar signal-to-noise ratio,  and in situations where the impact of ambient
conditions is relatively small and highly predictable. These conditions are
often met, for example, in space-based telescopes.  However,  the situation is
often different for the reduction of data from ground-based observatories.  The
reasons for this include the complexity of general purpose instruments that are
now routinely employed, the rapid upgrade pace necessary to exploit advances in
technology and science goals, and the variety of effects imposed by varying
atmospheric conditions. In many cases, supervision and interaction with the
data reduction process is, therefore, still  essential to obtain sufficiently
high quality results even from fairly routine observations. 

The general concept of astronomical data reduction that does not employ a fully
integrated pipeline has not substantially changed in the past decades.
Researchers organise their data, and use a mixture of general purpose and
highly specialized tools, inspecting the results of each step.  Such tools are
available in environments such as MIDAS \citep{bib:midas}, IRAF \citep{bib:iraf}
and IDL\footnote{IDL is a trademark of Research Systems Inc., registered in the
United States}, or as stand-alone programs.  What has changed is the number,
complexity and interdependence of steps needed to accomplish the data
reduction. In this situation, the efficiency of the data reduction process can
be vastly improved by automating the previously manual workflow of organising
data, running individual steps, and transferring results  to subsequent steps,
while still using the same routines to carry out individual reduction steps.  

The most commonly used approach to automate a data reduction workflow by
individual researchers is to employ a scripting language such
as Python \citep[e.g.][]{bib:fors}.  This approach works well with a relatively
small number of reduction steps, and in situations where the data organisation
and book-keeping are fairly simple. In more complex situations, such scripts
are themselves complex programs that cannot easily be modified. 

In this paper, we describe the usage of a general workflow engine to automate
the data reduction workflow for astronomical observations. While this approach
is relatively new for the field of astronomy
\citep[e.g.][]{bib:astrowork1,bib:astrowork2}, it has been widely used in other
fields of science including biology and chemistry \citep{bib:wf_bio},
meteorology \citep{bib:wf_meteo} and economics \citep{bib:wf_eco}.
For that reason, we  discuss  in detail the methods and functionalities that
are necessary to use such a system for astronomical data reduction, 
 and present ESO's new ``Recipe flexible execution workbench''
(\reflexstyle{Reflex}) environment as a specific implementation of such a
system.  

The structure of the paper is as follows. In Sec.~2,  we describe the main
principles and architecture of our design independent of a particular
implementation. In Sec.~3, we discuss how these principles can be implemented
in a specific workflow application, using our \reflexstyle{Reflex}
implementation as an example. Finally, in Sec.~4 we conclude with a discussion
of the impact of performing data reduction in this way.

%

\section{Architecture of astronomy data reduction workflows}

\subsection{Data organisation}\label{sec:organisation}

   \begin{figure} \includegraphics[width=\columnwidth]{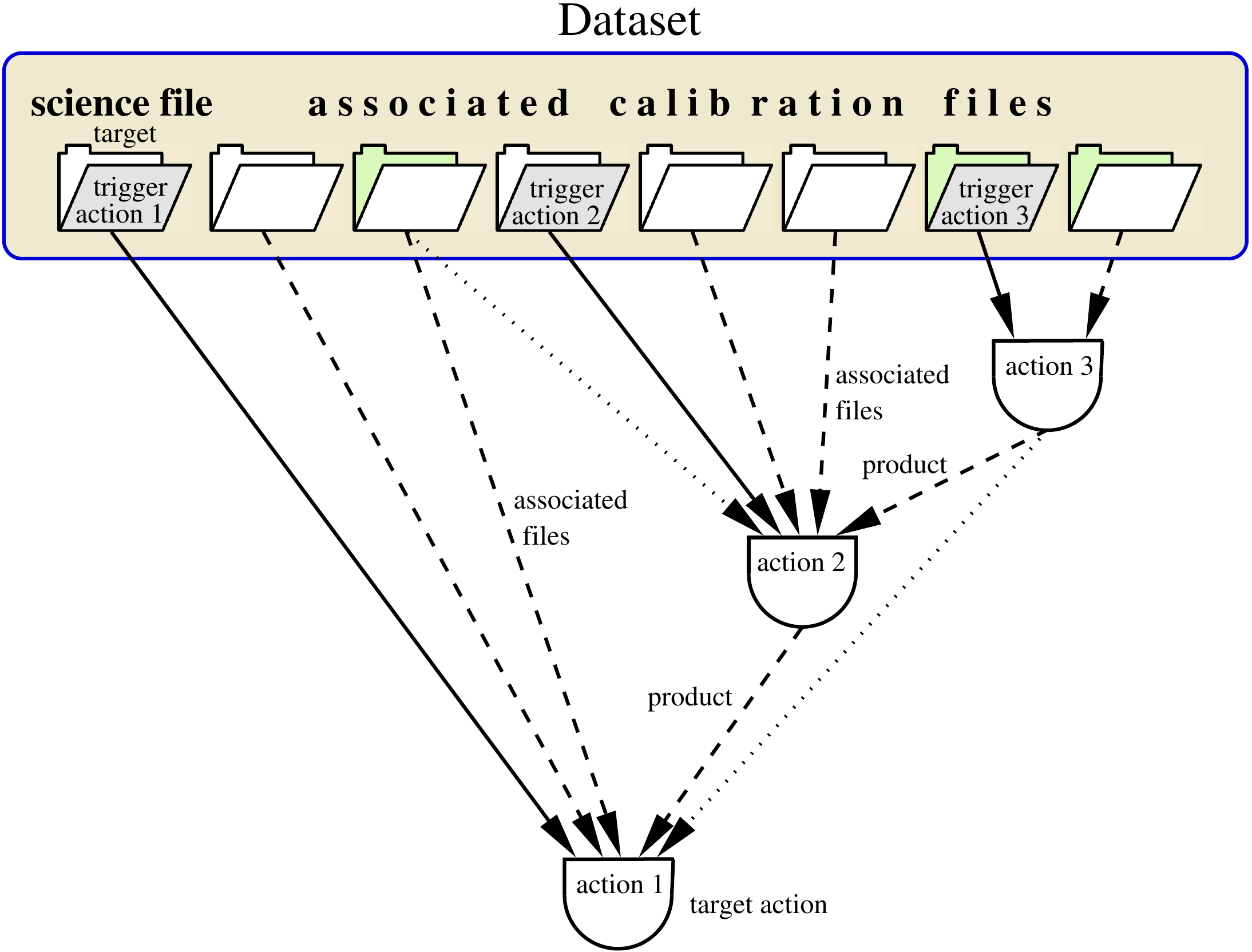}
           \caption{Example of a simple \reflexstyle{data set} and its
           organisation. The \reflexstyle{data set} contains all files
           necessary to produce the science data product of the workflow.  This
           includes the science associated calibration files.  These files
           are organised using a set of \reflexstyle{action}s that are shown as
           shield-shaped symbols.  The target files are directly connected to
           the target \reflexstyle{action} that is the root of the graph.
           Files that are connected to an \reflexstyle{action} with a solid
           line are the \reflexstyle{trigger} for that action.  Properties of
           the \reflexstyle{triggers}  are used to select associated files for
           an \reflexstyle{action}.  The associated files are connected to an
           \reflexstyle{action} with dashed or dotted lines. To highlight
           files that are connected to more than one \reflexstyle{action}, a
           dashed line is used for one of these connections, and a dotted line
           for the other one. The \reflexstyle{purpose} of a file is the
           connection between the file and the \reflexstyle{target action}.
           Symbols with tinted background indicate files that have multiple
           \reflexstyle{purpose}s, i.e. there are multiple paths from the file
           to the \reflexstyle{target action}.  } \label{fig:organigram}
   \end{figure}

Astronomical data consist of collections of files that include both the
recorded signal from extraterrestrial sources, and  metadata such as
instrumental, ambient, and atmospheric data.  Such a collection of files is the
raw output from one or several observing runs,  and consists of ``science
files'' that contain the primary science observations to be analysed.

In addition,  it might include files that are not
directly related to the current observations, such as calibration files that
are routinely collected for a given instrument.  Hereafter, we refer to the
input files for the data processing as ``raw files'', as opposed to files that
are created during the data processing and that we will refer to as
``products''.  We use the term ``calibration file'' for any raw file that is
not a science file.

In order to discuss data reduction in general terms, we introduce the following
terminology.  The goal of data reduction is to process sets of files, which we
refer to  as the \reflexstyle{target}s of a data reduction workflow. The result
of this processing is to create a \reflexstyle{target product}.  In most cases,
the \reflexstyle{targets} of a data reduction workflow will be the science
files, and the \reflexstyle{target product} is then the science data product to
be used for scientific analysis.  The \reflexstyle{target} files can be
naturally grouped into sets that are reduced together. Such a group of
\reflexstyle{target} files, together with other files needed to process them,
is referred to as a \reflexstyle{data set}.  A \reflexstyle{data set} is
``complete'' when it contains all necessary files to reduce the
\reflexstyle{target}s, and ``incomplete'' if some of those files are missing.
\label{sec:complete}

``Data organisation'' is the process of selecting \reflexstyle{data set}s from
a  larger collection of files, and recording information on the type of files
and the reasons for selecting them. This larger collection of files might
be the result of a pre-selection process that assures that low quality or
defective data are not considered at this stage.  Organising data  is a
complex and time-intensive procedure that is typically among the first tasks of
a data reduction workflow \cite[e.g.][]{bib:vvds}. Hereafter, we will refer to
the whole data reduction workflow including data organisation simply as a
``workflow'', whereas we will use the term ``data processing workflow'' for the
processing of data that follows the data organisation.

The first step in data organisation is to classify files, i.e. to determine the
data content of each file from its metadata. The goal of classification is to
assign a \reflexstyle{category} to each file. An example of such a
\reflexstyle{category} is ``flatfield for filter I''.  The next step is to
identify the \reflexstyle{target}s, and group them into \reflexstyle{data set}s
that are incomplete at this stage.  Subsequently, calibrations are added to the
\reflexstyle{data set}s.  Calibration files for each \reflexstyle{data set} are
selected by analysing the metadata of the \reflexstyle{target}s and that of
other available files that potentially qualify for inclusion in a
\reflexstyle{data set}.

This cascade of selection criteria naturally maps into a data graph
\label{sec:datagraph} as illustrated in Fig.~\ref{fig:organigram}.  The links
between elements of the graph show the flow of metadata that originates from
the raw files.  The graph is directed, i.e. links between elements have a
direction to distinguish between incoming and outgoing information.  The nodes
of the graph define necessary procedural steps in the assembly of a
\reflexstyle{data set}, and we refer to them as \reflexstyle{action}s.  The
targets of the workflow connect directly to the root node (action~1) that is
therefore called the \reflexstyle{target action}. 

Each \reflexstyle{action} has several incoming files connected to it.  Some of
those are used to define selection properties of other input files to that
\reflexstyle{action}. For example, an \reflexstyle{action} might specify to
select flat field images  that use the same filter as the science image.  We
use the notation that the files that are used to define properties of other
files, in our example the science files,  are the \reflexstyle{trigger} for
that \reflexstyle{action}, and their links are shown as solid lines in
Fig.~\ref{fig:organigram}.  The \reflexstyle{trigger} of the
\reflexstyle{target action} are the \reflexstyle{target}s of the workflow.

All \reflexstyle{action}s other than the \reflexstyle{target action} have one
or several outgoing links that connect them to subsequent
\reflexstyle{action}s.  These outgoing links pass on  metadata  that are
extracted from the input files to the next \reflexstyle{actor}.  They are
therefore called \reflexstyle{product}s of an \reflexstyle{action}.  These
\reflexstyle{products} do not necessarily correspond to actual physical
products produced during data reduction, and the actual physical products
created during data reduction do not necessarily appear in the data
organisation graph. Instead, the products in the data organisation graph  are
used as a logical scheme to define the selection of data. For example, for the
purpose of data organisation, it  is not necessary to define a
\reflexstyle{target product}, even when the data processing workflow creates
one. This is because  the nature and properties of the target product have no
impact on the data selection.

Each raw file is the origin of at least one path along the direction of the
links that lead to the \reflexstyle{target action}.  This reflects the fact
that \reflexstyle{data set}s  only include raw files that are needed to process
the targets of the workflow.  A  path runs either directly from the files to
the target action, or passes through other actions on its way. We refer to such
a path as one of the \reflexstyle{purpose}s of a file.  The
\reflexstyle{purpose}s of a file are important information for the data
processing (see Sec.~\ref{sec:purpose}).

\label{sec:ocaexample} In Fig.~\ref{fig:organigram2}, we show the data graph
for a specific example with the same symbols used in Fig.~\ref{fig:organigram}.
The simple example is  an image that needs a bias frame, a flatfield and a dark
frame for its processing. The flatfield needs to be taken with the same optical
filter as the science frame, whereas the dark frame needs to be taken with the
same exposure time as the science frame.  Therefore, flatfields  and dark
frames with these properties must be identified among available files, and one
of each must be selected according to criteria such as the closeness of the
time of observation to that of  the science frame.  After this step, more
calibration files need to be added to the \reflexstyle{data set} that are used
to reduce the calibration files.  The selection criteria for those files depend
on properties of calibration files instead of the \reflexstyle{target}s of the
workflow. In the current example, the flatfield itself needs a dark frame for
its processing, and this dark frame needs to match the exposure time of the
flatfield, not that of the science frame.  The science frame, flatfield frame
and dark frame in turn might all require their own bias frame for reduction.
The \reflexstyle{actions} in this case are given specific labels, namely
``proc\_dark'', ``proc\_flat'' and ``proc\_image''. Note that the action
``proc\_dark'' is shown twice, reflecting the fact that it is used twice, once
to select darks for the flatfield, and a second time to select darks for the
image.

We note that the  topology of the graph might differ between \reflexstyle{data
set}s  even for the same kind of data. For example, in one \reflexstyle{data
set} the input dark frames for the science and flat frames might be identical,
in another one they might differ.  The task of data organisation is to create
such a graph for each individual data set. 

   \begin{figure} \includegraphics[width=\columnwidth]{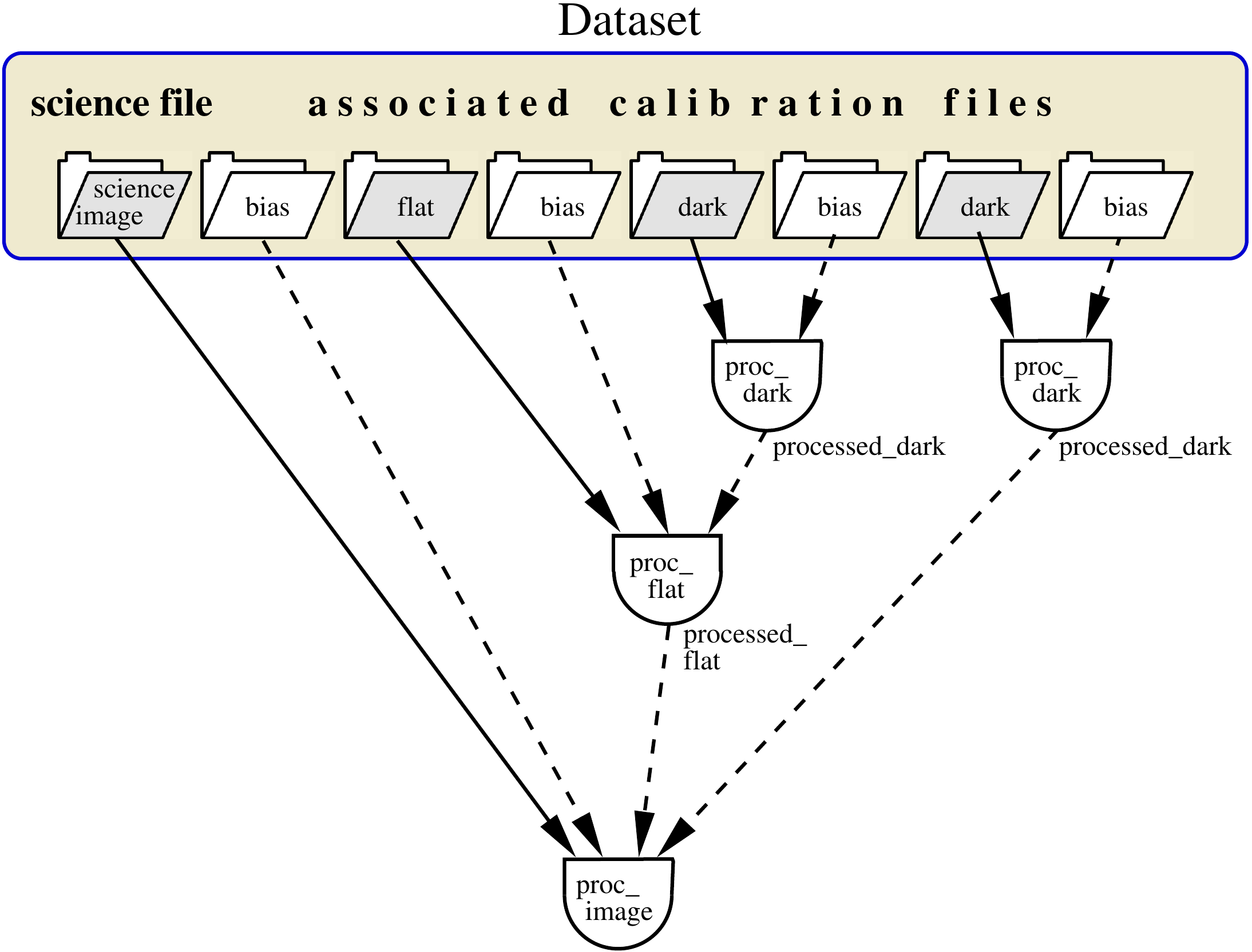}
           \caption{Data graph for a data set to process images as
           described in the text. The symbols used are the same as in
           Fig.~\ref{fig:organigram}.  In the case shown here, each file has a
           unique \reflexstyle{purpose}, and therefore no dotted lines are
           used. The \reflexstyle{action} ``proc\_dark'' is used to select 
           different darks for the flat frame and the science image. Therefore,
           it appears twice in the graph.
             } \label{fig:organigram2} \end{figure}

\subsection{Design of data organisation and data processing systems}

The data organisation discussed in Sec.~\ref{sec:organisation} and the  data
processing that follows the data organisation both describe relations among
different \reflexstyle{categories} of data.  These relations are
interdependent, in the sense that a change in the selection of data might
require some change in the data processing, and vice versa.  The question
therefore arises as to whether the best architecture is to derive these
relations from a common source, or whether the information recorded in these
relations is sufficiently different to warrant their independent
implementation. 

In general, a specific selection of data does not uniquely specify the data
processing sequence. Very  different data processing workflows can be
constructed to use a given selection of data.  Only the most basic data
processing follows the data organisation process one-to-one, but this case is
rarely used in practice. The data processing part of the  workflows are, in
general, more complex than the data organisation, and are also more frequently
subject to change and optimization during the data reduction process. The
\reflexstyle{purpose} of a file records an aspect of the selection criteria
used to include this file in a \reflexstyle{data set}. It is up to the workflow
design to decide how this information is used.

For example, a \reflexstyle{category} of files (such as a flatfield) might be
selected to match the date of the science frames and is, therefore, assigned a
corresponding  \reflexstyle{purpose}. This does not necessarily mean that 
these flatfields are  exclusively used to flatfield the science frames, but the
data processing workflow might also use them to  flatfield standard star flux
calibration data. For spectroscopy, it is not always clear whether the best
flatfields for the flux calibrator are those that are taken close in  time to
the target spectrum or those taken close in time to the flux calibrator.  This
decision depends on a complex set of circumstances.  A workflow might include
conditional and/or interactive parts to help the user make that decision. 

Another difference between data organisation and data processing is
that, while some steps in the data processing are closely related to a
specific selection of data, others are completely independent of it. For
example, a step that only modifies intermediate products has no impact on the
data selection or organisation. Steps that make small adjustments to
intermediate products are often added or removed during data reduction.  Any
system that mixes the data selection and data processing workflows is then,
necessarily, much more complex than either of the two components individually.
One design goal for a workflow system is to make modification of  the data
reduction as simple as possible.  This is helped by clearly separating the data
organisation from the data reduction steps. 

We therefore advocate a design that not only separates the implementation of
the two steps, but also  uses a different methodology to define the two tasks.
Each of them should be geared towards the specific needs of each step.  The
data organisation is usually closely related to the instrument properties, the
observing strategy and the calibration plan. The strategy for data organisation
therefore rarely changes after the observations have taken place. Interactivity
in that part will create overheads that do not outweigh the expected benefits.
On the other hand, the data processing is, in general, highly interactive and
experimental, and the final strategy is rarely known at the time of
observation.  The best values for data reduction   parameters and even the
chosen strategy might depend on the properties of individual \reflexstyle{data
set}s.

An efficient way to implement a data organisation is, therefore, a rule-based
system that can accommodate complex, instrument-specific rules and can be run
to organise either locally stored data or data extracted from an archive
repository using pre-defined rules.  The syntax of the  rules must be able to
describe the method of creating  data graphs such as the ones discussed in
Sec.~\ref{sec:datagraph}.  Such  data organisation is particularly efficient if
it is carried out by raw-data archives that have any potentially useful
calibration file available for retrieval. For example,  ESO offers an archive
service
``calselector''\footnote{\url{http://www.eso.org/sci/archive/calselectorInfo.html}}
that selects, organises and provides access to data in a manner similar to the
one described above.

In contrast, data processing after the data organisation    benefits from
interactive, graphical and dynamic elements. An efficient way to provide this
is to use a  workflow application that allows the implementation of workflows
that can be easily modified for experimentation and optimization during data
processing. It is important that these interactive features can be turned off
once a workflow has been tuned and optimized, in order to allow time-intensive
processing to be carried out in a batch mode.

\subsection{Functionalities of a rule-based data organiser}

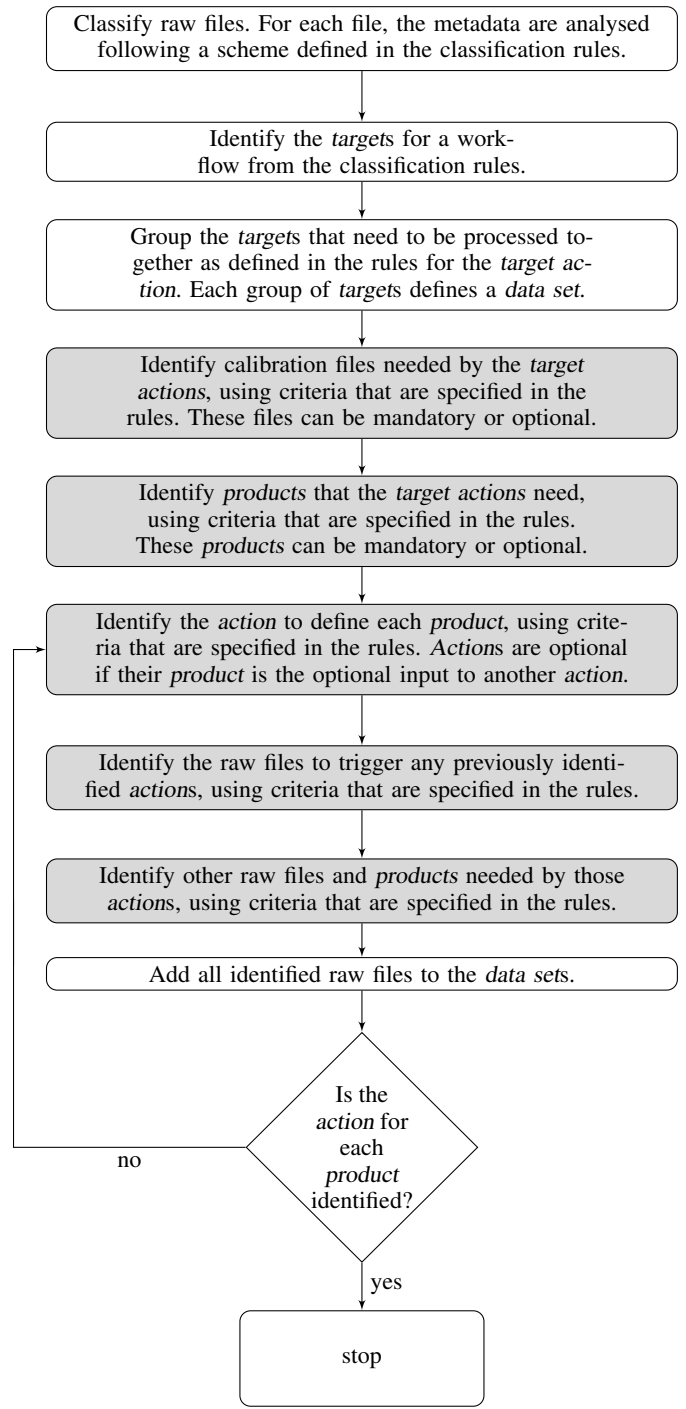
\begin{figure}

\tikzstyle{decision} = [diamond, draw, 
    text width=4.5em, text badly centered, node distance=3cm, inner sep=0pt]
\tikzstyle{block} = [rectangle, draw,
    text width=0.9\columnwidth, text centered, rounded corners, minimum height=1em]
\tikzstyle{filledblock} = [rectangle, draw, fill=black!15,
    text width=0.9\columnwidth, text centered, rounded corners, minimum height=1em]
\tikzstyle{smallblock} = [rectangle, draw, 
    text width=3cm, text centered, rounded corners, minimum height=4em]
\tikzstyle{line} = [draw, -latex']
\tikzstyle{lineb} = [draw]
\tikzstyle{dummy} = []
    
\centering
\begin{tikzpicture}[node distance = 1.5cm, auto]

\node [block] (0) {Classify raw files. For each file, the metadata are analysed
following a scheme defined in the classification rules.};

\node [block, below of=0, node distance=1.5cm] (1) { Identify the
\reflexstyle{target}s for a workflow from the classification rules.};

\node [block, below of=1, node distance=1.5cm] (2) { Group the
\reflexstyle{target}s that need to be processed together as defined in the
rules for the \reflexstyle{target action}. Each group of \reflexstyle{target}s
defines a \reflexstyle{data set}.};

\node [filledblock, below of=2, node distance=1.7cm] (3) { Identify calibration
files needed by the  \reflexstyle{target actions}, using criteria that are
specified in the rules.  These files can be mandatory or optional.};

\node [filledblock, below of=3, node distance=1.7cm] (4) { Identify
\reflexstyle{products} that the \reflexstyle{target actions} need, using
criteria that are specified in the rules.  These \reflexstyle{products} can be
mandatory or optional.};

\node [filledblock, below of=4, node distance=1.7cm] (5) {  Identify the
\reflexstyle{action} to define each \reflexstyle{product}, using criteria that
are specified in the rules.  \reflexstyle{Action}s are optional if their
\reflexstyle{product} is the optional input to another \reflexstyle{action}. };
        
\node [filledblock, below of=5, node distance=1.7cm] (6) { Identify the raw
files to  trigger any previously identified \reflexstyle{action}s, using
criteria that are specified in the rules.};

\node [filledblock, below of=6, node distance=1.5cm] (7) { Identify other  raw
files and \reflexstyle{products} needed by those \reflexstyle{action}s, using
criteria that are specified in the rules.};

\node [block, below of=7, node distance=1.1cm] (8) { Add all identified raw
files to the \reflexstyle{data set}s. };

\node [dummy, left of=7, node distance=0.51\columnwidth] (d1) {};

\node [dummy, left of=8, node distance=0.51\columnwidth] (d2) {};

\node [decision, below of=8, node distance=2.3cm] (decide) { Is the
\reflexstyle{action} for each \reflexstyle{product} identified? };

\node [smallblock, below of=decide, node distance=2.8cm] (stop) {stop};
\path [line] (0) -- (1);
\path [line] (1) -- (2);
\path [line] (2) -- (3);
\path [line] (3) -- (4);
\path [line] (4) -- (5);
\path [line] (5) -- (6);
\path [line] (6) -- (7);
\path [line] (7) -- (8);
\path [line] (8) -- (decide);
\path [lineb] (decide) -| node [near start] {no} (d1);
\path [line] (d2) |- (5);
\path [line] (decide) -- node {yes}(stop);
\end{tikzpicture}

\caption{High-level flow chart of a data organiser. If any step in a shaded box
fails for any given \reflexstyle{data set}, then this \reflexstyle{data set} is
marked as ``incomplete''.  }\label{fig:flowchart} 
\end{figure}

A software program that uses rules to organise data  as advocated above can
produce the data graph discussed in Sec.~\ref{sec:organisation} by a set of
steps shown in the flow chart of Fig.~\ref{fig:flowchart}.  The output of the
data organisation is a list of \reflexstyle{data set}s. A \reflexstyle{data
set} is marked as ``complete'' if there are files that satisfy the criteria
used in steps shown in  shaded boxes in Fig.~\ref{fig:flowchart}. It is marked
as ``incomplete'' if any one of those criteria are not satisfied by any
existing file.  Each file in the output \reflexstyle{data sets} is described by
the file name, the \reflexstyle{category} of the file as defined in the rules,
and the \reflexstyle{purpose} of the file. The \reflexstyle{purpose} of the
file is recorded as the concatenation of the names of the \reflexstyle{action}s
that link the file to the \reflexstyle{target action}. In the  example
discussed in Sec.~\ref{sec:ocaexample}, the flatfield is selected based on
properties (in this case the filter) of the science frame.  The selection rules
are defined in the \reflexstyle{action} called ``proc\_science''.Then the bias
frame is selected for this flatfield based on properties of the flatfield (e.g.
observing date or read-out mode). This selection is defined in the action
``proc\_flat''. The \reflexstyle{purpose} of this bias frame, as well as the
flatfield, is then ``proc\_science:proc\_flat'', while the bias frame that
matches the properties of the science frame, as well as the science frames
themselves, have the \reflexstyle{purpose} ``proc\_science''. The other biases
in this example have the purpose ``proc\_science:proc\_flat:proc\_dark'' and
``proc\_dark''. The different biases have different \reflexstyle{purpose}s so
that the workflow can process them separately.  \label{sec:purpose} A given
file might have several different \reflexstyle{purpose}s if it is selected
multiple times by the rules (see Fig.~\ref{fig:organigram}).  An example of
this is when the same bias frame matches the selection rules for both the
flatfield and the science frames.

%

\subsection{Data processing workflows}

There are different ways to carry out the task of reducing astronomical data,
even when the applications used for individual reduction steps are fixed.  One
approach is to sort data by \reflexstyle{category}, and process each
\reflexstyle{category} in sequence. For example, one might start by processing
all the bias frames for all \reflexstyle{data set}s as the very first step,
then proceed to subtract combined biases from all relevant data, and continue
with, for example,  producing flatfields. 

A different approach is to fully process a single \reflexstyle{data set},
performing all necessary steps to see the final result for the first
\reflexstyle{data set} in the shortest possible time. Each intermediate
product, such as a combined bias, is produced only when it is needed.

The former approach has the advantage that it simplifies book-keeping, in that
the only necessary initial sorting is by file type. Operations of the same kind
are all performed together. The parameters for every task are optimized by
repeatedly inspecting results. Once a good set of parameters is found, it is
applied to all files of the same kind. This approach is efficient in the sense
that identical operations are carried out only once, while the effort for
bookkeeping is minimal.  It is, therefore, often used when workflows are
manually executed by scientists that call individual steps in sequence and
book-keeping is carried out ad-hoc without software tools. 

The advantage of the latter approach is that it allows for easier inspection of
the impact of any change in parameters or procedures on the quality of the
final \reflexstyle{target product} of a workflow. This is particularly
important when data reduction strategies are still experimental and being
tested. This approach also delivers the results faster in that it  only
executes the steps that are needed for a given \reflexstyle{data set} and
thereby more quickly produces the \reflexstyle{target product} for the first
\reflexstyle{data set}. 

The advantages of both of these approaches can be combined with the following
design. As in the second approach, data are processed one \reflexstyle{data
set} at a time. Data reduction steps that need to be executed several times
with different input files from the same \reflexstyle{data set} are carried out
in succession.  For example, the step to combine bias frames is executed for
the biases to be applied to the science frame, and immediately afterwards the
bias frames to debias the flatflields are processed, and so on. The inputs and
outputs of each individual data reduction step are stored in a database for
re-use later.  Whenever a reduction step is called, this database is checked
for previous calls to the reduction step with the same input files and
parameters.  If such a previous call exists, then the reduction step is not
executed and instead the previous results are re-used. We call the
feature to re-use products created by previous executions of a procedure the
``lazy mode'' \label{sec:lazy}. There might be cases when such a  re-use of
products is not desired. A lazy mode should, therefore, always be an option of
each individual step in a workflow.  An example for the efficiency gain from
using  this mode is  a set of combined biases that are used by the science and
flatfield frames of a \reflexstyle{data set}, and in addition by the
calibration frame of another \reflexstyle{data set}. The combination of biases
is carried out only once, and is used in three different places.  One advantage
of our approach is that it is as efficient as the first of the above
approaches, but produces the science results quickly and provides the user
experience of the second approach.  Another advantage is that subsequent runs
of the workflow can use this database of intermediate products to redo the
reduction with changed parameters in a very efficient manner. If a parameter or
input file for any step changes, then the result for this step will change. The
change in one of the intermediate products might require the re-execution of
some but not all of the subsequent steps. The database can be used to
automatically identify products that can be re-used from previous runs, and the
steps that need to be repeated.

\setlength{\fboxsep}{0.2mm} 
 
 \begin{figure} \includegraphics[width=\columnwidth]{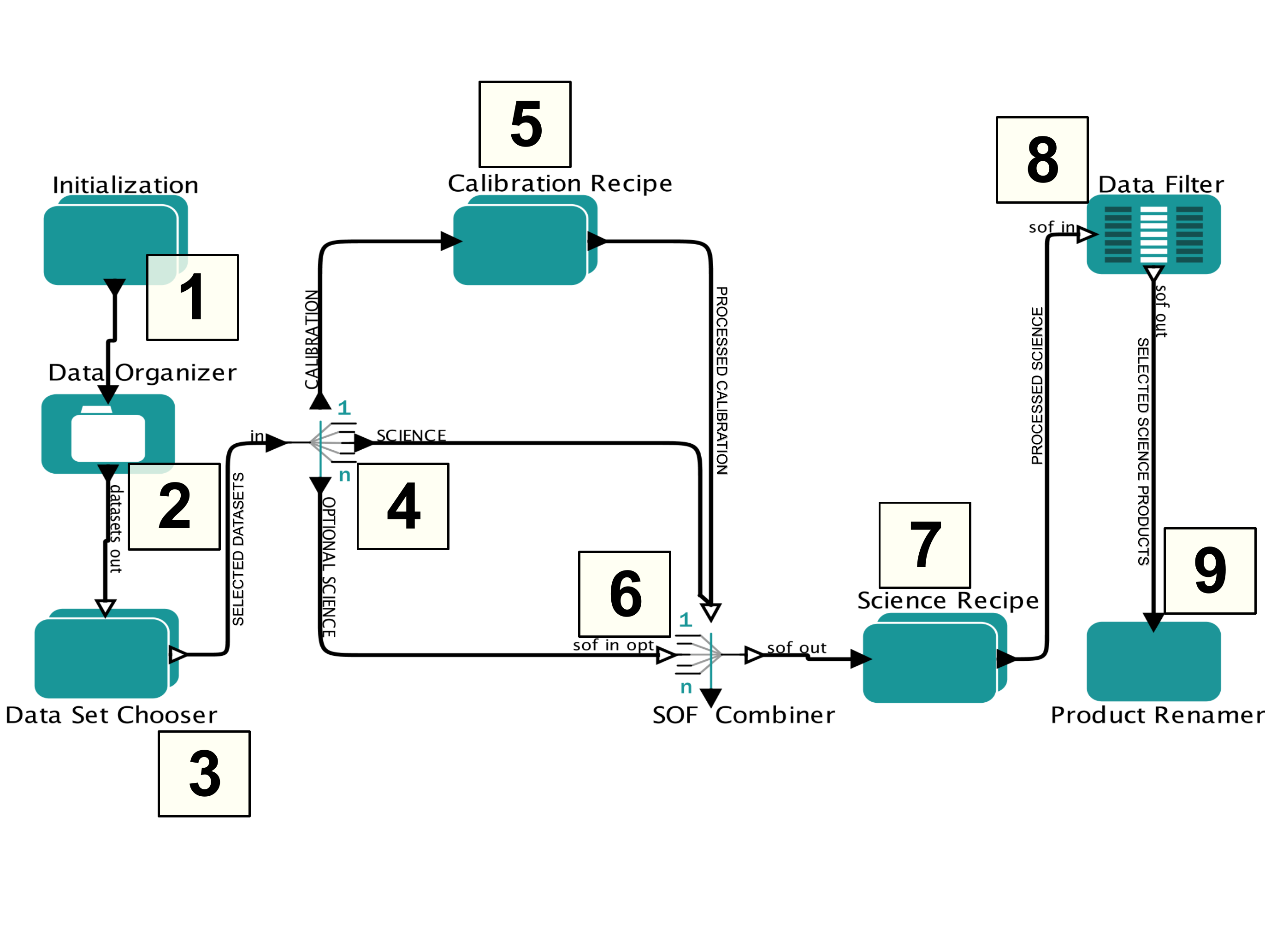}
         \caption{Example of a basic \reflexstyle{Reflex} workflow. The figure
         uses the graphical elements of a Kepler workflow
         (Sec.~\ref{sec:kepler}). The lines indicate the flow of files and are
         labelled by their contents. The ``optional science'' files are files
         that are used to process the science data, but the processing can
         proceed even if they are not available (see Sec.~\ref{sec:optional}).
         The workflow includes two data processing steps, one for calibration
         and one for science processing (labelled \frame{ 5} and \frame{ 7},
         respectively).  The elements of the workflow are:  an initialization
         \frame{ 1} that sends the input directories to the data organiser, the
         data organiser \frame{ 2}, a data set chooser \frame{ 3} that allows
         interactive selection of a \reflexstyle{data set},  the file router
         \frame{ 4}  that directs different \reflexstyle{categories} of files
         to their destinations, a SOFCombiner \frame{ 6} that bundles the input
         for the science step, and a data filter and product renamer (\frame{
         8} and \frame{ 9}, respectively) that organise the output products
         from the workflow.} \label{fig:basicworkflow} \end{figure}

The implementation of this workflow design requires three levels of grouping of
data.  A schematic diagram of such a workflow is shown in
Fig.~\ref{fig:basicworkflow}.  The highest level of grouping are the
\reflexstyle{data set}s, as discussed in Sec.~\ref{sec:organisation}.  This
task is carried out by a data organiser (step~\framebox{\tiny 2} in
Fig.~\ref{fig:basicworkflow}). Subsequently, the files in each
\reflexstyle{data set} are sorted by \reflexstyle{category} and directed to the
reduction steps that need this particular \reflexstyle{category} of files, a
step that is performed by a file router  (step~\framebox{\tiny 4}).  This is the
level that describes the data reduction strategy and is shown in the design of
a workflow.  Each reduction step might be called repeatedly with different
input files. For that purpose, a third level of grouping is needed to group
files that are processed together with separate calls of the reduction step.
This is part of the functionality of the reductions steps \framebox{\tiny 5} and
\framebox{\tiny 7}.

\section{Implementation}

While the principles discussed in this paper do not depend on a specific
software implementation, it is useful to discuss them in the context of, and
with the terminology used, in a specific environment.  Several software
environments to design and execute workflows exist
\cite[e.g.][]{bib:otherworkflow}. For the \reflexstyle{Reflex} project, we
evaluated and partially implemented some of the concepts discussed in this
paper in several open source workflow engines.  In the end, we decided to use
the Kepler \label{sec:kepler} workflow application \citep{bib:kepler} to
implement \reflexstyle{Reflex}, because of its large suite of  available
components, and its robust support for conditional branching, looping, and
progress monitoring.  In this section, we introduce the terminology and
summarize the most important features of Kepler.  For more details, see the
Kepler User
Manual\footnote{\url{https://code.kepler-project.org/code/kepler-docs/trunk/outreach/documentation/shipping/2.4/UserManual.pdf}}.


\subsection{The Kepler workflow engine}

A workflow application is a software system designed to run sequences of
stand-alone programs, where the programs depend on the results  of each
other. Components of a workflow are representations of these programs, as
well as elements that manage the results and communication between them.
 In Kepler, components of the workflow are called
``actors''. In the graphical interface, actors are represented by green boxes
(see Fig.~\ref{fig:keplerlayout}). Actors have named input and output  ``ports'' to
communicate with each other. The communication is implemented by exchanging
objects called ``tokens'' that travel along connections between the output port
of one actor to the input port of another actor.  These  connections  are
called ``relations'' and are represented by lines. Output ports emit tokens
after the execution of an actor is finished, and input ports consume tokens when
execution of the actor starts. The availability of tokens is a crucial factor
in determining the sequence of triggering the actors. 

   \begin{figure*}
      \includegraphics[width=\textwidth]{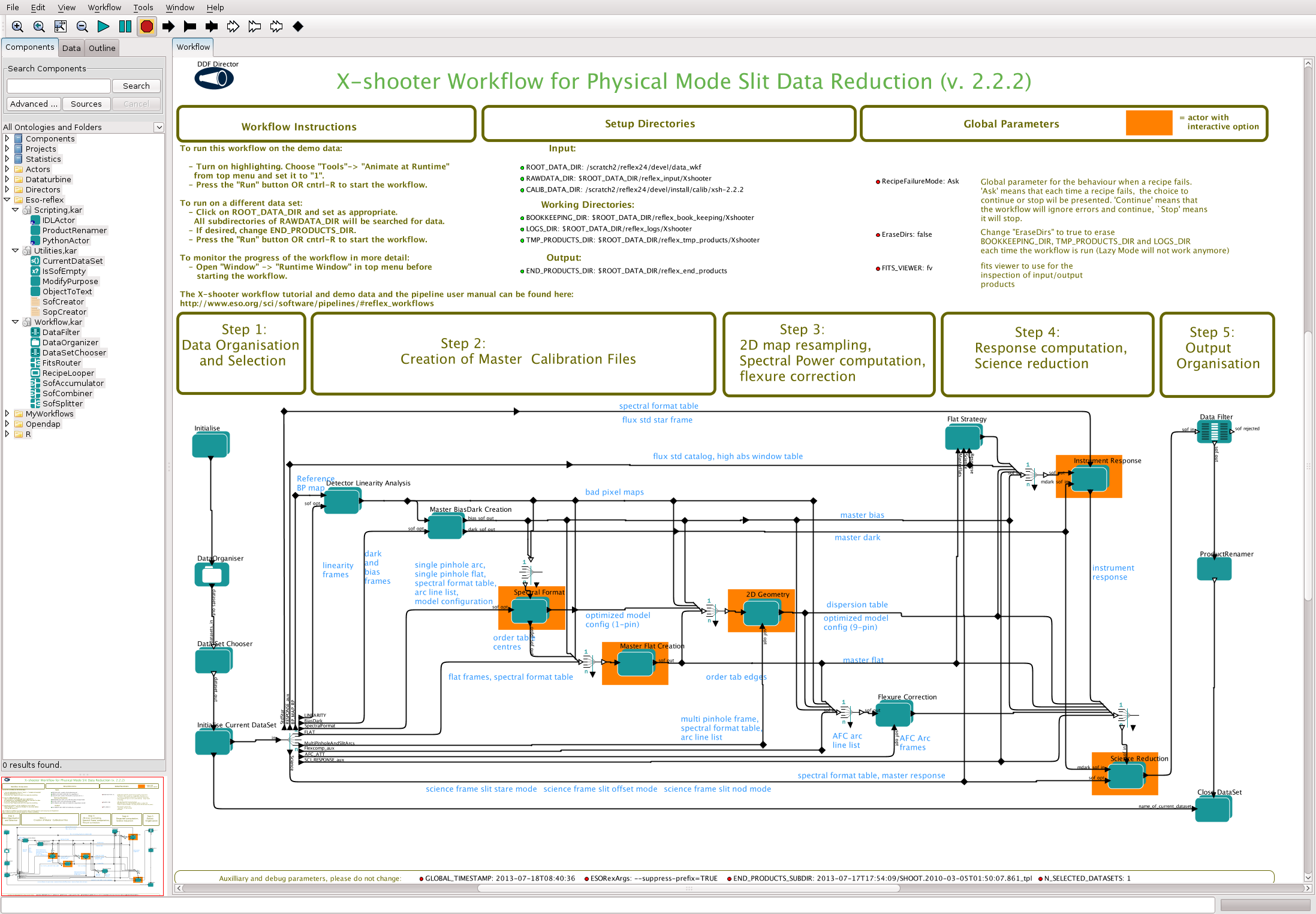} \caption{The Kepler user
      interface loaded with the   \reflexstyle{Reflex} workflow for ESO's
      X-Shooter instrument. The top section defines the input directories and
      user preferences.  It is usually sufficient to specify the raw data
      directory to run the workflow on a new data set.  The execution of a
      workflow is started with the run button in the top left panel.  The
      workflow includes 8 recipe executers that run the recipes  necessary to
      reduce X-Shooter data.  Actors with an orange background include
      interactive steps that can display the result of the recipe, and allow for
      the optimization recipe settings. The workflow includes  an specially implemented
      actor called  ``Flat Strategy'' that is specific to the X-Shooter workflow. 
      This actor allows the user to select a flatfielding
      strategy. Depending on the chosen strategy, files will be routed differently. 
        } \label{fig:keplerlayout}
      \end{figure*}

The \reflexstyle{relation}s between actors themselves do not define the
temporal sequence of the execution of actors. A scheduler is required to
trigger the execution of each actor.  A scheduler in a Kepler workflow is
called a ``director''. The terminology  of Kepler follows the metaphor of film
making, where a director instructs actors that carry out their parts.
\reflexstyle{Reflex} uses the ``Dynamic Data Flow'' (DDF) director that allows
the workflow execution to depend on the results of actors and supports looping
and iterating. The basic algorithm used by the DDF director is to repeatedly
scan all actors and identify those that can be executed because all of the
necessary input is available.  It then selects one of the actors for execution
based on minimizing unused tokens and memory.  The details are extensively
discussed in \cite{bib:ddf}.  It should be noted that an actor of a workflow
can itself be a sub-workflow.  Such ``composite actors'' might include their
own directors.

The Kepler workflow application provides a graphical interface to create,
edit  and execute workflows.  A large number of general purpose actors are
bundled with the environment. There are several ways to monitor the progress of
a workflow, pause it, or stop it. 

\begin{table*}
\caption{OCA rules syntax}             
\label{tab:oca}      
\centering                          
\begin{tabular}{l l l}        
\hline\hline                 
rule type & syntax   \\    
\hline                \\        
Classification& \tt if {\sl condition} then \{ REFLEX.CATG = "{\sl category"}; REFLEX.TARGET=``[T,F]'' \} \\
\\
Organisation & 
\begin{minipage}[t]{0.74\paperwidth}%
     \tt   minRet = {\sl i};  \\
\tt select execute({\sl actionname}) from inputFiles where {\sl conditions} [group by {\sl keyword-list}]\\
\end{minipage} \\

Association& 
\begin{minipage}[t]{0.74\paperwidth}%
\tt action {\sl actionname}
\{ \\
\phantom{xxx} minRet = {\sl i}; maxRet = {\sl j};  \\
\phantom{xxx} select files as {\sl label} from inputFiles where  {\sl conditions} \\
\phantom{xxxxxxxxxxxxxxxxxxxxxxxxxxxxxxxxxxxxxxxxxxxxxxxxxx}[closest by {\sl keyword}]; \\
\phantom{xxx} product {\sl label} \{ REFLEX.CATG = {\sl category} \}; \\
  \} \\
\end{minipage}
\\
\hline                                   
\end{tabular}
\tablefoot{The table lists a simplified version of the  OCA rules syntax appropriate for data organisation in \reflexstyle{Reflex} workflows. The {\sl conditions} define \reflexstyle{categories} of  FITS files by their header keywords and may include logical and arithmetic expressions. The \reflexstyle{label}s in the association rule are used for logging purposes and 
are usually set to the \reflexstyle{category} of the file defined in the rule.}
\end{table*}

\subsection{The \reflexstyle{Reflex} Environment}

We have produced the software package \reflexstyle{Reflex} to implement the
design discussed in this paper using Kepler workflows.  \reflexstyle{Reflex}
consists of a collection of actors that support the execution of astronomical
applications.  A shared characteristic of commonly used astronomical
applications is that they read data and metadata from FITS files, are
configurable with parameters, and produce output FITS files called products.
\reflexstyle{Reflex} supports any application of this kind that can be started
from a command line.  Hereafter, we refer to such applications as ``recipes''.
The primary task of \reflexstyle{Reflex} is to route the necessary  input files
to the recipes. This includes both files in a \reflexstyle{data set} and files
created during execution of the data processing workflow. In addition,
\reflexstyle{Reflex} is able to create and send lists of parameters to recipes.

To achieve these tasks, \reflexstyle{Reflex} uses  two kinds of objects called
``set of files'' (hereafter SOF) and  ``set of parameters'' (hereafter SOP).
These objects are used as tokens in a workflow.  A SOF contains a list of
files. The record for each file consists of the file name, the checksum, the
\reflexstyle{category} and a list of \reflexstyle{purpose}s for that file. A
SOP contains a list of parameters, and the record for each parameter consists
of its name and value. \reflexstyle{Reflex} actors use and process these
objects. 

The construction of an input SOF, to be fed to a recipe, needs to consider the
\reflexstyle{category} and the \reflexstyle{purpose}  of a file.  For every
file in a \reflexstyle{data set}, these file properties are determined during
file organisation according to the pre-defined rules.  For products, these
properties have to be determined during the execution of the workflow. It is
important to note that these two file properties are handled differently. Every
recipe needs to be aware of the file \reflexstyle{category} of its input files.
For example, a recipe that combines flatfields might use dome flat exposures
and bias frames as input files, and these files need to be identified to the
recipe. The mechanism to identify files to the recipes is different in
different environments. For example, IRAF uses different input
parameters for different file types, whereas ESO's CPL recipes \citep{bib:cpl}
use text files with file tags to identify the file types. In both cases,
\reflexstyle{Reflex} uses the \reflexstyle{category} to identify these file
types.  \reflexstyle{Reflex} workflows therefore need to explicitly use the
exact names known to the recipes for its \reflexstyle{categories}.  The data
organisation  rules have to generate these exact names.  In contrast, recipes
are oblivious to the \reflexstyle{purpose} of a file. The recipe to combine
flat field frames does not need to know how and where the combined flatfields
will be used. Therefore, the processing of the \reflexstyle{purpose} is
completely handled by workflow actors. 

As discussed in Sec.~\ref{sec:purpose}, a \reflexstyle{purpose} is a
concatenation of \reflexstyle{action}s used to organise the input data. The
name of an \reflexstyle{action} is arbitrary and, therefore, it is never used
explicitly in the workflow. Instead, \reflexstyle{Reflex} uses the overriding
principle that recipes receive files of the same, but arbitrary
\reflexstyle{purpose}. Actors compare and manipulate, but never decode the
\reflexstyle{purpose}. There are three standard operations on the
\reflexstyle{purpose}, they are called \reflexstyle{pass-through},
\reflexstyle{set-to-universal} and \reflexstyle{trim}. The operation
\reflexstyle{pass-through} simply reads the \reflexstyle{purpose} of a file,
and passes it on without any modification. The operation
\reflexstyle{set-to-universal} replaces an existing \reflexstyle{purpose} with
a new one with the protected name  \reflexstyle{universal}.  The
\reflexstyle{universal} \reflexstyle{purpose} is a wildcard that may be
substituted by any other defined \reflexstyle{purpose} depending on the
circumstances.  The operation \reflexstyle{trim}
modifies a \reflexstyle{purpose} by removing the last \reflexstyle{action} from
a \reflexstyle{purpose} that consists of at least two concatenated
\reflexstyle{action}s, and sets a \reflexstyle{purpose} that consists of a
single \reflexstyle{action} to \reflexstyle{universal}.  Workflows are designed
so that the input to any recipe consists of files with a single identical
\reflexstyle{purpose}. These  operations are sufficient to design workflows
that collect all necessary input files for each recipe, by selecting all files
with identical \reflexstyle{purpose}s to be processed. 

\label{sec:operations} The usage of these operations can best be explained with
examples. The most commonly used operation is \reflexstyle{trim}. In
Sec.~\ref{sec:purpose}, we already used the example of bias frames with the
\reflexstyle{purpose} ``proc\_science:proc\_flat'' and ``proc\_science'', and
flatfield  and science frames with the \reflexstyle{purpose} ``proc\_science''.
When the flatfield is to be processed, the workflow selects the file with the
\reflexstyle{category} ``flat'', and all files with identical
\reflexstyle{purpose}. In our example, these are the bias and flatfield files
with \reflexstyle{purpose} ``proc\_science:proc\_flatfield''.  The output
product, i.e. the processed flatfield, should be assigned the
\reflexstyle{trim}med input \reflexstyle{purpose}. In our case, the
\reflexstyle{purpose} ``proc\_science:proc\_flat'' is reduced to
``proc\_science''.  In a subsequent step, a science recipe collects all the
files with purpose ``proc\_science'' for the input.  This will include the
processed flatfield file, the bias frame selected to match the properties of
the science frame, and the science frame itself. 

The operation \reflexstyle{pass-through} is used for recipes that only use
intermediate products as inputs.  If such a recipe is needed in the chain (e.g.
to  smooth the flatfields in the above example), this recipe 
should pass-through the \reflexstyle{purpose} of its input file to the product file, so that the
\reflexstyle{purpose} of the smoothed flatfields is still ``proc\_science''. In
general, any recipe that has no impact on the data selection should pass-through
the \reflexstyle{purpose} of the input files.  

Finally, the \reflexstyle{set-to-universal} operation can be used for files
with a unique \reflexstyle{category} that can be processed independently of their
usage in the workflow.  For example, a bad pixel map that is used by many
different recipes in a workflow can be given the \reflexstyle{purpose}
``universal'' to simplify the workflow design.

These three operations allow an efficient and elegant assembly of input files
for recipes. Different  operations might be necessary under special
circumstances and a flexible system will allow these to be  implemented.  An
important design principle for any operation on the \reflexstyle{purpose} of a
file  is that it should never explicitly use the name of the
\reflexstyle{purpose}. The names assigned by the rules are arbitrary, and a
change of those names should not impact the workflow execution. 

\subsection{\reflexstyle{Reflex} actors}

In order to implement these principles, \reflexstyle{Reflex} provides 17
essential actors. A complete list of actors is given in Appendix~A and
are described in detail  by \cite{bib:reflex_manual}.  The actors can be grouped
into the data organiser, actors to process and direct tokens, actors to
execute data reduction recipes written in one of several supported languages, and
actors that provide interactive steps in a workflow. In this section, we discuss
these features and options to illustrate how the principles laid out above can
be implemented in concrete software modules. 

\begin{table*}
\label{tab:ocaexample}      
\centering                          
\caption{Simple Example of OCA rules}             

\hrulefill

\begin{center}
Classification rules
\end{center}
\begin{verbatim}
if TYPE=="OBJECT" then {REFLEX.CATG = "science_image";  REFLEX.TARGET="T";}
if TYPE=="FLAT"  then {REFLEX.CATG = "flat";}
if TYPE=="DARK"  then {REFLEX.CATG = "dark";}
if TYPE=="CALIB" and EXPTIME==0 then {REFLEX.CATG = "bias";}
\end{verbatim}
\begin{center}
Organisation rules
\end{center}

\begin{verbatim}
select execute(proc_dark) from inputFiles where REFLEX.CATG=="dark" 
select execute(proc_flat) from inputFiles where REFLEX.CATG=="flat" 
select execute(proc_image) from inputFiles where REFLEX.CATG=="science_image" 
\end{verbatim}
\begin{center}
Association rules
\end{center}

\begin{verbatim}
action proc_dark
{
  select files as bias from inputFiles where REFLEX.CATG=="bias" ;
  product processed_dark { REFLEX.CATG="processed_dark";}
}

action proc_flat
{
  select files as bias from inputFiles where REFLEX.CATG=="bias" ;
  select files as processed_dark from inputFiles where REFLEX.CATG=="processed_dark"
                                             and inputFile.EXPTIME==EXPTIME;
  product processed_flat { REFLEX.CATG="processed_flat";}
}

action proc_image
{
  select files as bias from inputFiles where REFLEX.CATG=="bias" ;
  select files as processed_dark from inputFiles where REFLEX.CATG=="processed_dark"
                                                       and inputFile.EXPTIME==EXPTIME;
  select files as processed_flat from inputFiles where REFLEX.CATG=="processed_flat" 
                                                       and inputFile.FILTER=FILTER ;
}
\end{verbatim}
        
\hrulefill
\tablefoot{The table lists an example of a set of OCA rules that can produce the 
data organisation shown in Fig.~\ref{fig:organigram2}. }
\end{table*}

\subsubsection{Data organiser and rule syntax}

\label{sec:do} For \reflexstyle{Reflex}, we opted to implement a program
DataOrganiser that carries out the organisation of local data fully
automatically using a set of user-supplied, human-readable rules. The input of
the DataOrganiser is a set of FITS files and the classification rules; the
output is a collection of \reflexstyle{data set}s. The rules are based on the
principles discussed above.  It should be re-iterated that, while definitions
of \reflexstyle{action}s could be used to define a data structure in sufficient
detail to allow automatic derivation of a simple data processing workflow, this
is not the approach that we adopt here.  Instead, the rules are used only to
organise the data, while the workflow to reduce the data is not constrained to
using the selected data in any particular manner.

The DataOrganiser is the first   actor after the initialization  in any
\reflexstyle{Reflex} workflow. It  organises  the input FITS files  according
to the workflow-specific input rules, and the output are \reflexstyle{data
set}s that are either marked complete or incomplete (see
Sec.~\ref{sec:complete}). The execution of a \reflexstyle{Reflex} workflow is
triggered by sending an input token to the DataOrganiser actor.

The DataOrganiser recognizes rules that use the syntax of a special language
called OCA. The OCA language has been developed at ESO \citep{bib:ocarules} and
is designed to describe the {\bf O}rganisation, {\bf C}lassification, and
{\bf A}ssociation of FITS files based on their FITS header keywords
\citep{bib:fits}.  OCA is used for multiple purposes within ESO's data flow
system \citep{bib:dataflow}, and interpreters are embedded in a number of
applications.  Therefore, OCA rules to organise data are available for most
instruments on ESO's telescopes.  The details of the language are described in
\cite{bib:ocaman}. The language has all of the features needed to define rules
for data organisation. Here, we summarize a subset of the OCA language that is
useful for the data organisation discussed in this paper.

The OCA language recognizes three types of rules. They are:
\begin{enumerate}  

\item Classification Rules. Classification rules  define file
        \reflexstyle{categories} based on any logical combination of conditions
        on FITS keywords.  The syntax of the classification rules is given in
        row 1 of Tab.~\ref{tab:oca}.  The classification defines the keyword
        REFLEX.CATG as the  \reflexstyle{category} of the file,
        and this keyword can be used like any other FITS keyword in the header
        by other rules.  A simple example for the usage of a classification
        rule is to assign to a file the \reflexstyle{category} ``bias'' if the
        header keyword ``EXPTIME'' is set to the value ``0''.  The
        classification rules also define whether a set of files is the target
        of the workflow or not.

\item Organisation Rules. Organisation rules define \reflexstyle{action}s and
        the groups of files that \reflexstyle{trigger} them.  The rules define
        a name for each \reflexstyle{action}  so that it can be referred to by
        other rules.  The syntax of the organisation rules is given in row 2 of
        Tab.~\ref{tab:oca}.  The rules include an optional specification of the
        minimum number of files needed to \reflexstyle{trigger} the
        \reflexstyle{action}.  This minimum number is used to determine whether
        a \reflexstyle{data set} is complete or not. There is no maximum number
        because there are no defined criteria to select among files that match
        the condition.  A simple example is to group at least 3 dome flat
        frames by filter, and \reflexstyle{trigger} an \reflexstyle{action}
        called ``proc\_flat'' that combines flatfields.

\item Association Rules. Association rules define  ``associated files'', i.e.
        input files and \reflexstyle{product}s that are needed by an
        \reflexstyle{action} in addition to the \reflexstyle{trigger}.  The
        syntax of the association rules is given in row 3 of
        Tab.~\ref{tab:oca}.  There is an unlimited number of ``select''
        statements that define conditions to select files. In addition to the
        conditions, a ``closest by'' statement can be used to select those
        files that have a value for a given keyword that is as  close as
        possible to that of the \reflexstyle{trigger}. If there is no ``closest
        by'' statement, then the time of observation will be used to select
        among several files that satisfy the conditions.    Each select
        statement can be preceded by an optional specification of the minimum
        and maximum number of files needed for each \reflexstyle{category}.
        This mechanism allows to define optional input files that are not
        essential for a workflow but that will be  used if present. The
        association rules also define names for \reflexstyle{categories} of
        \reflexstyle{products} that can be referred to by other rules.  A
        simple example is that the proc\_flat \reflexstyle{action} needs a
        combined bias frame and produces a \reflexstyle{product}
        ``MasterFlat''.  

\end{enumerate}

These rules are sufficient to describe the data
graphs discussed in Sec.~\ref{sec:datagraph} and shown
in Fig.~\ref{fig:organigram} and~\ref{fig:organigram2}.  In
Tab.~\ref{tab:ocaexample}, we show as a specific example the rules that
describe the data organisation for an image that needs a flatfield and
a dark frame for its processing, as discussed in Sec.~\ref{sec:ocaexample} and
shown in Fig.~\ref{fig:organigram2}.  The first block classifies available
files as ``science\_image'', ``flat'' or ``dark'' based on the header keyword
``TYPE'', and as ``bias'' based on the fact that the  value of the header
keyword ``EXPTIME'' is 0.  The next three ``select'' statements define the
three \reflexstyle{actions} ``proc\_dark'', ``proc\_flat'' and ``proc\_image'',
and their \reflexstyle{triggers} ``flat'', ``dark'' and ``science\_image''.
What follows are the association rules that specify that the
\reflexstyle{action} ``proc\_dark'' needs a bias as input and outputs a product
called ``processed\_dark'', the \reflexstyle{action} ``proc\_flat'' needs this
``processed\_dark'' and a bias, and outputs a ``processed\_flat''.  Finally,
the action ``proc\_image'' needs a ``bias'', the ``processed\_dark''
and the ``processed\_flat''. The association rules also specify that darks are
selected to match the exposure time of the ``science\_image'' or the ``flat'',
and flats are selected to match the filter of the ``science\_image''. The
application of these rules can lead to a data set organized as shown in
Fig.~\ref{fig:organigram2}.  However, the same rules can also lead to a data
graph with different topology for a different data set.  For example, if both
the science image and the flat have the same exposure time, the application of
the rules might select the same dark frame for both the flat and the science
image. The power of the data organiser is to use such abstract rules to select
optimal data sets based on the metadata of the available files.

\subsubsection{Actors for data processing}

The purpose of a data processing workflow is to execute a series of recipes.
The recipes can be written in any language, but must accept the basic input,
and provide the basic output information, needed to run the workflow.  In
particular, the recipes must accept FITS files that are categorized, and
generate products as FITS files and the information to categorize them.
\reflexstyle{Reflex} provides three actors to execute recipes. They are called
``PythonActor'', ``IDLActor'' and ``RecipeExecuter''.  The PythonActor is used
to run Python scripts that, in turn, can call, for example, shell commands,
IRAF tasks via the PyRAF interface \citep{bib:pyraf}, or MIDAS programs via the
pyMIDAS interface \citep{bib:pymidas}. The IDLActor is used to run IDL programs, and the
RecipeExecuter executes CPL recipes.
The basic function of all
three actors is to filter and send the files of the input SOF to the recipe,
and create and emit the  output SOF with the products of the recipe.  The
\reflexstyle{purpose} of the product files is constructed from those of the
input files using one of the standard operations described in
Sec.~\ref{sec:operations}. All CPL recipes can be queried to report their input
and output in a well-defined format. This feature  is used by
\reflexstyle{Reflex} to automatically generate parameter lists and ports. For
Python and IDL, simple interfaces are provided that can be added to any
program.

   \begin{figure}
      \includegraphics[width=\columnwidth]{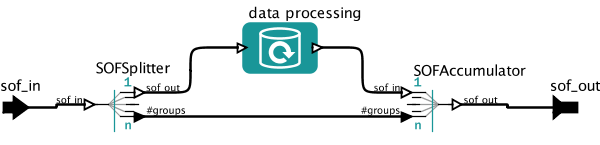}
      \caption{A data processing actor embedded in a SOFSplitter and SOFAccumulator to manage
      repetitive executions for files with different \reflexstyle{purpose}s. }
      \label{fig:recipe_loop}
   \end{figure}

\subsubsection{Actors for file routing}

The top level task when designing a data processing workflow is to decide on the
cascade of file processing, i.e. the routing of files by
\reflexstyle{category}. In \reflexstyle{Reflex}, users of a workflow are
presented with a visual diagram that shows the directional flow of files with
different \reflexstyle{categories} to the corresponding data processing actors
(see Fig.~\ref{fig:basicworkflow}). The \reflexstyle{data set}s created by the
DataOrganiser are SOFs that contain a full set of file
\reflexstyle{categories}. An actor is needed to direct the different
\reflexstyle{categories} of files in a \reflexstyle{data set} to the respective
data processing actors. In \reflexstyle{Reflex}, this actor is called the
FitsRouter. It takes a single \reflexstyle{data set} SOF as input, and creates
SOFs that contain input files selected by \reflexstyle{category}  from the
\reflexstyle{data set}.  Different output SOFs are emitted from separate ports,
that are connected to data processing actors. For each output port, one, or
several, file \reflexstyle{categories} sent to this port are explicitly
specified by name.  The primary use of the FitsRouter is to  select the
\reflexstyle{categories} of raw files  in a \reflexstyle{data set} that are
needed for each data processing actor,   whereas products needed as input
arrive directly from the data processing actor via dedicated
\reflexstyle{relation}s.

The routing by \reflexstyle{category} assures that  a recipe receives all
necessary file \reflexstyle{categories}  at the time that it is executed. If
there are files with \reflexstyle{categories} that are not needed by the
recipes, they can be filtered out by the data processing actor. What remains is
the task to select among all of the files of a given \reflexstyle{category}
those that should be processed together by the recipe.  In
\reflexstyle{Reflex}, this is implemented as an actor ``SOFCombiner'' that
bundles the different input files for a recipe into a single SOF.  The
SOFCombiner has two input ports, one is for mandatory files and another one for
optional files.\label{sec:optional} Both of them are multiple ports, i.e.
several \reflexstyle{relation}s can be connected to either input port. The
tokens sent via different \reflexstyle{relations} to a multiple port are in
different \reflexstyle{channels} of the port. The SOFCombiner creates a single
output SOF that includes input files selected by \reflexstyle{purpose}. The
selection rule is that  only files with a \reflexstyle{purpose} that is present
at each of the input \reflexstyle{channels}, at the mandatory input port, are
passed.  The desired selection of all files with the appropriate
\reflexstyle{purpose} is achieved when at least one of the input channels
includes only files that are the necessary input for the recipe, typically the
\reflexstyle{trigger} for the recipe.   All other channels can include any
file, and the SOFCombiner automatically selects the correct input for the
recipe.  The algorithm used by the SOFCombiner  uses comparison of
\reflexstyle{purpose}s as the only method and consists of the following two
steps.

\begin{enumerate}

        \item Find \reflexstyle{purpose}s that are present at each input
                \reflexstyle{channel} of the mandatory input.  A
                \reflexstyle{universal} \reflexstyle{purpose} counts as a match
                to any other \reflexstyle{purpose}.  \label{sec:universal}

        \item Send all files, both from the mandatory and optional port, that
                match  any \reflexstyle{purpose} found in step 1 to the output
                SOF.  Again, a \reflexstyle{universal} \reflexstyle{purpose}
                counts as a match to any other \reflexstyle{purpose}.
\end{enumerate}

 This simple but powerful algorithm assures that the files of
 the same \reflexstyle{purpose} in the output SOF are necessary and sufficient
 to run the intended data processing recipe. This fact is then used by a
 combination of two  actors called  ``SOFSplitter'' and ``SOFAccumulator''. The
 former  splits an input SOF by \reflexstyle{purpose} and emits a separate SOF
 for each \reflexstyle{purpose}.  The latter collects several SOFs  that arrive
 at the same port and combines them into a single SOF.  These actors are used in
 combination with, and always bracket, a data processing actor (see
 Fig.~\ref{fig:recipe_loop}).  The net effect of this combination is that the
 recipes called by the data processing actor are executed multiple times, and
 the result is a single SOF that includes all of the products.  This SOF can
 then be used by the next SOFCombiner to select the files for the next data
 processing step.

The algorithms discussed above are an elegant and efficient way to implement
the most common routing needs without repetition of information. In addition,
explicit operations  on file properties can be used to implement special needs.
A common application is conditional routing of files, i.e. workflows in which
files are routed differently depending on some data properties or user choices.
Kepler provides a large number of general purpose actors to implement
conditional and/or iterative branches in a workflow, and  stand-alone actors, to
manipulate the \reflexstyle{category} or \reflexstyle{purpose} of a file, are
either provided by \reflexstyle{Reflex} or can easily be implemented (e.g. as a
Python script). As emphasized earlier, in any manipulation of the
\reflexstyle{purpose}, the \reflexstyle{purpose} should never explicitly be
called by name within the workflow to avoid unnecessary dependencies of the
workflow on syntax choices in the rules.  For example, the case discussed above
that a flatfield file is selected to be taken close in time to   the science
spectrum, but is used to flatfield the flux calibration file can easily be
implemented with such a customized \reflexstyle{purpose} processing
script. 

   \begin{figure}
      \includegraphics[width=\columnwidth]{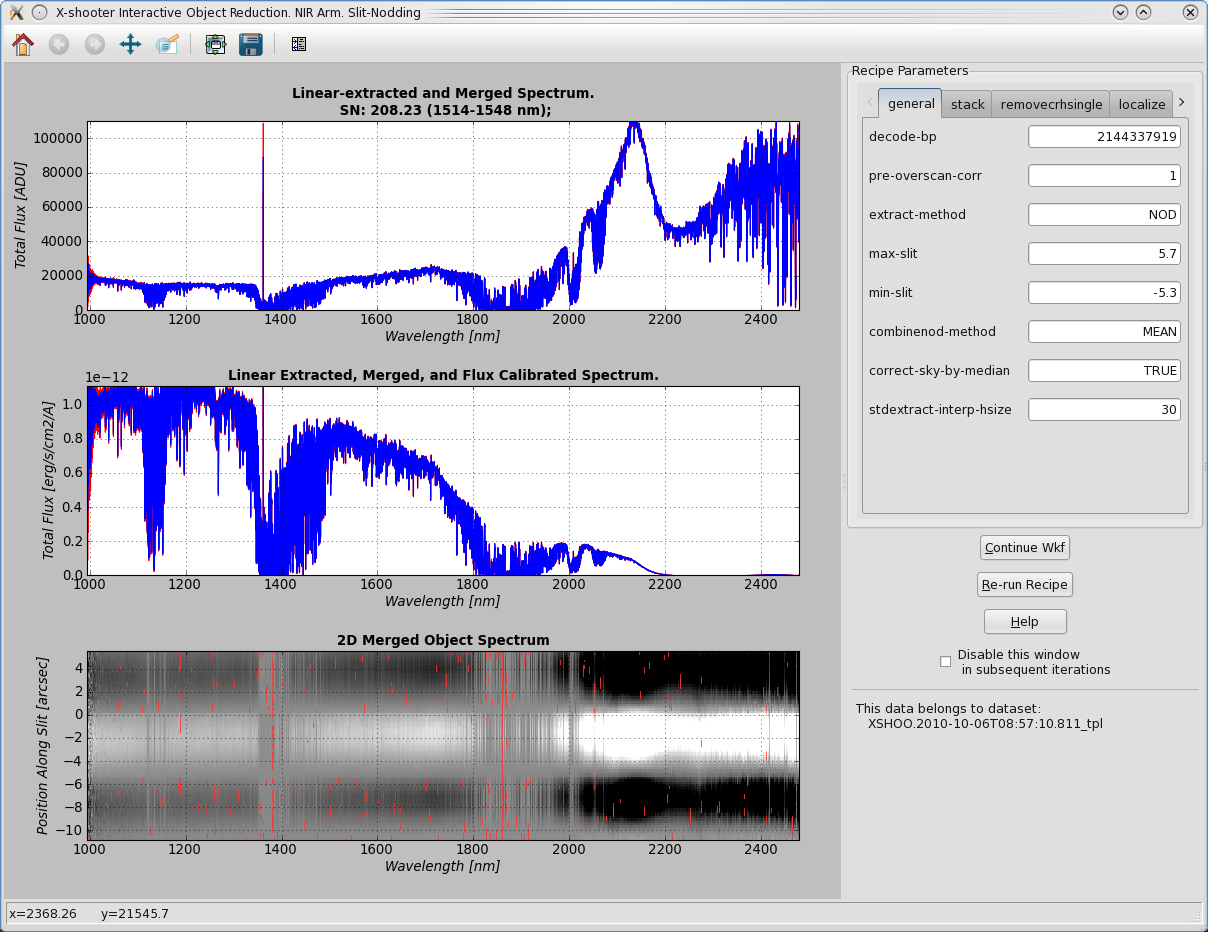}
      \caption{Example of an interactive interface created with the \reflexstyle{Reflex} Python library. 
      The plots on the left hand side can be interactively manipulated to inspect the data products
      in different ways. The panel on the right hand side allows  the user to modify recipes parameters, 
      and re-execute the recipe or continue with the workflow.}
      \label{fig:interactive}
   \end{figure}

\subsubsection{Interactive actors}

One reason why automated workflows are an efficient way of data reduction is
that the user can intercept the processing at any stage and interact with the
workflow.  A major contribution to the interactive user experience comes from
the workflow application that provides tools to monitor, pause  and modify the
workflow itself.  Additional tools are needed to provide application-specific
ways to inspect and influence the execution of the workflow.
\reflexstyle{Reflex} provides several interactive actors, and a Python library
to implement actors that can create customized plots and
allow recipe parameters to be modified during the execution of a workflow.  An
example of an interface created with this library is shown in
Fig.~\ref{fig:interactive}. Ready-to-use interactive actors that have been
developed for \reflexstyle{Reflex} include the DataSetChooser to interactively
inspect and select \reflexstyle{data set}s, the DataFilter to inspect and
filter SOFs, and the ProvenanceExplorer to inspect the provenance of a product and
its history from repeated runs of a workflow. All interactive actors and
features can easily be turned off when starting \reflexstyle{Reflex}, to allow
a workflow to be run in batch mode once it has been adapted and optimized.


\subsection{Modularity of \reflexstyle{Reflex} workflows}

Kepler provides an easy way to create modular workflows. A composite actor is
an actor that itself contains a workflow, and composite actors can be nested to
arbitrary depth. Placing each data processing actor, together with its
supporting actors, into a composite actor leads to a clean and intuitive view
of the whole data processing workflow. 

The layout of a workflow, whether it is modular or not, does not uniquely
define a sequence of actor executions. For example,  a scheduler might decide
to alternate processing of actors contained in different composite actors.
However, workflow execution is more intuitive when each composite actor is
completed before processing proceeds to other actors.  In Kepler, this can be
achieved by placing an  appropriately configured director into each composite
actor.

%

\subsection{Book-keeping and product organisation}

The efficiency of the workflow execution relies on rigorous book-keeping that
stores relevant information in a database for easy retrieval and processing.
During execution of the workflow, the input and output files of each step in
the workflow, as well as all parameters used for the processing are stored in
the database. For each file, the file name and the  checksum are recorded.

The two main uses of the database are the implementation of the lazy mode
described in Sec.~\ref{sec:lazy}, i.e. the keeping track of products for later
re-usage, and the organisation of the output files in a user-friendly way. For
the lazy mode, checksums and creation dates can be used to detect changes in
input files of the same name.  The main output of a workflow are the files
produced by the final science processing step, i.e.  the science data products
of a workflow.  Intermediate products produced by previous steps are often
needed to evaluate the science data products, troubleshoot, or investigate the
optimization of the products.  For that purpose, each science data product
should be associated with the input data and parameters  used to generate it.
The input files might themselves include products from previous steps, that are
associated to the input of that step. At the conclusion of a workflow, all
files used and produced during its execution can be organised in a directory
tree that can be browsed either with a specialized tool or with a file browser.

\section{Summary and conclusions}

In this paper, we describe how a workflow application can be used to
automate an astronomical data processing workflow. We propose a specific
design for such a workflow, and present the application  \reflexstyle{Reflex} that implements
this design within the Kepler workflow engine. The key advantages of
automated workflows over alternative methods such as scripting or monolithic
data processing programs are the built-in tools for progress
monitoring, and the ability to modify the data organisation and data flow
efficiently. 

The specific advantages of our \reflexstyle{Reflex} implementation include: 

   \begin{enumerate}

      \item Selecting and  organising the input data is a significant task for
              any astronomical data reduction. A rule-based data organiser is
              used  to formalize the selection criteria and to fully automate
              the organisation of data. The automated data organisation can
              be followed by an interactive step to inspect and modify the
              chosen \reflexstyle{data set}s.

      \item \reflexstyle{Reflex} allows users to monitor the progress of data
              reduction, interact with the process when necessary, and  modify
              the workflow. A graphical user interface can be used to develop
              and experiment with workflows. At the same time, workflows can be
              executed in a completely non-interactive batch mode to allow
              processing of large data sets and/or for computational time-intensive
              processing. 

      \item Re-reduction, after a change in input files or parameters, is
              efficiently carried out by only re-running those steps that are
              affected by this change. A modern reduction process might use
              hundreds of files with dozens of different
              \reflexstyle{categories}, and any number of data reduction steps.
              Changing a single parameter in one of the steps or switching a
              single input file might trigger a complex cascade of necessary
              re-runs of steps.  Recognizing those steps and re-executing them
              is fully automated in \reflexstyle{Reflex}.

   \end{enumerate}

The execution time of the data organiser strongly depends on the
complexity of the rules, the total number of files, and the files in each
category.  For data from a typical observing run, the first time data
organization might take on the order of a minute on a typical desktop
workstation. In subsequent runs the lazy mode will reduce this time by a very
large factor. The execution of the data processing workflow itself adds a
fraction of a second to the stand-alone execution of each recipe. The default
memory allocation for \reflexstyle{Reflex} is 1536 MB in addition to the memory
requirement of the recipes. This allocation can be reduced for simple workflows
if necessary.

So far, \reflexstyle{Reflex} workflows have been developed for the most
commonly used instruments on ESO's Very Large Telescope (VLT), namely FORS2,
SINFONI, UVES, VIMOS, and X-Shooter, as well as the newly commissioned KMOS.
They are distributed to users to provide them with a pre-packaged workflow that
works out-of-the-box to reduce VLT data.  All ESO Reflex workflows are
intuitive to understand, as each includes a detailed tutorial and a
comprehensive demonstration data set.   As such, even novice users can easily
modify and experiment with the workflows.  \reflexstyle{Reflex} workflows are
bundled with the corresponding instrument pipelines and the
\reflexstyle{Reflex} environment. The whole package can be installed with a
single installation script available at
\url{http://www.eso.org/reflex/}.  ESO expects to develop
\reflexstyle{Reflex} workflows for all future VLT instruments.

\bigskip \begin{acknowledgements} The Kepler software is developed and
        maintained by the cross-project Kepler collaboration, which is led by a
        team consisting of several of the key institutions that originated the
        project: UC Davis, UC Santa Barbara, and UC San Diego. We acknowledge
        useful discussions with Reinhard Hanuschik on some concepts discussed in
        this paper.  
\end{acknowledgements}


\begin{appendix}

        \section{List of \reflexstyle{Reflex} actors}

A complete description of Reflex actors is given by \citet{bib:reflex_manual}. Here, we list the standard Reflex actors in alphabetical order.

\begin{itemize}

        \item{{\bf{DataFilter:}}} Interactive actor to inspect and select FITS
                files.
        \item{{\bf{DataOrganiser:}}}
                Implementation of the rule-based data organiser as described in text.
        \item{{\bf{DataSetChooser:}}}
                Interactive actor to inspect files in a \reflexstyle{data set}, edit the selection,  and select \reflexstyle{data set}s to be reduced.
        \item{{\bf{FitsRouter:}}}
                Actor to route files by \reflexstyle{category}.
        \item{{\bf{IDLActor:}}}
                Interface to configure and execute IDL scripts. 
        \item{{\bf{IsSofEmpty:}}}
                Actor that checks whether an SOF contains files. This actor is used to implement different data flows depending on the availability of some data.
        \item{{\bf{ObjectToText:}}}
                Actor to present \reflexstyle{Reflex} tokens in human readable form.
        \item{{\bf{ProductRenamer:}}} 
                Actor for renaming FITS files based on keywords of the file.
        \item{{\bf{ProvenanceExplorer:}}}
                Interactive actor to inspect products produced during a current or previous run of the workflow.
        \item{{\bf{PythonActor:}}}
                Interface to configure and execute Python scripts. 
        \item{{\bf{RecipeExecuter:}}}
                Interface to configure and execute CPL recipes.
        \item{{\bf{RecipeLooper:}}}
                Actor to implement looping over one or several recipes.

        \item{{\bf{SofCreator:}}}
                Actor to create a \reflexstyle{Reflex} Set of Files (SOF) token from a directory with files.
        \item{{\bf{SopCreator:}}}
                Actor to create \reflexstyle{Reflex} Set of Parameter (SOP) tokens.
        \item{{\bf{SOFAccumulator:}}}
                Actor to create a single SOF out of several input SOFs that arrive in sequence.
        \item{{\bf{SOFCombiner:}}}
                Actor to create a single SOF out of several SOFs that are available simultaneously.
        \item{{\bf{SOFSplitter:}}}
                Actor to split an SOF by file \reflexstyle{category}.
\end{itemize}

\end{appendix}

\end{document}